\documentclass[
]{ceurart}

\usepackage{listings}
\begin{document}

\copyrightyear{2023}
\copyrightclause{Copyright for this paper by its authors.
  Use permitted under Creative Commons License Attribution 4.0
  International (CC BY 4.0).}

\conference{HCMIR23: 2nd Workshop on Human-Centric Music Information Research, November 10th, 2023, Milan, Italy}

\title{The First Cadenza Signal Processing Challenge: Improving Music for Those With a Hearing Loss}

\author[1]{Gerardo Roa\ Dabike}[orcid=0000-0001-7839-8061]
\author[2]{Scott Bannister}[orcid=0000-0003-4905-0511]
\author[3]{Jennifer Firth}[orcid=0000-0002-7825-0945]
\author[1]{Simone Graetzer}[orcid=0000-0003-1446-5637]
\author[1]{Rebecca Vos}[orcid=0000-0002-2629-6271]
\author[3]{Michael A. Akeroyd}[orcid=0000-0002-7182-9209]
\author[4]{Jon Barker}[orcid=0000-0002-1684-5660]
\author[1]{Trevor J. Cox}[orcid=0000-0002-4075-7564]
\author[1]{Bruno Fazenda}[orcid=0000-0002-3912-0582]
\author[2]{Alinka Greasley}[orcid=0000-0001-6262-2655]
\author[3]{William Whitmer}[orcid=0000-0001-8618-6851]

\address[1]{Acoustics Research Centre, University of Salford, UK}
\address[2]{School of Music, University of Leeds, UK}
\address[3]{Hearing Sciences, School of Medicine, University of Nottingham, UK}
\address[4]{Department of Computer Science, University of Sheffield, UK}


\begin{abstract}
The Cadenza project aims to improve the audio quality of music for those who have a hearing loss. This is being done through a series of signal processing challenges, to foster better and more inclusive technologies. In the first round, two common listening scenarios are considered: listening to music over headphones, and with a hearing aid in a car. The first scenario is cast as a demixing-remixing problem, where the music is decomposed into vocals, bass, drums and other components. These can then be intelligently remixed in a personalized way, to increase the audio quality for a person who has a hearing loss. In the second scenario, music is coming from car loudspeakers, and the music has to be enhanced to overcome the masking effect of the car noise. This is done by taking into account the music, the hearing ability of the listener, the hearing aid and the speed of the car. The audio quality of the submissions will be evaluated using the Hearing Aid Audio Quality Index (HAAQI) for objective assessment and by a panel of people with hearing loss for subjective evaluation.
\end{abstract}

\begin{keywords}
  hearing loss \sep
  hearing aids \sep
  inclusive music \sep
  music quality \sep
  machine learning \sep
  signal processing \sep
  challenge
\end{keywords}

\maketitle

\section{Introduction}


There are many causes of hearing loss, including congenital hearing loss, chronic middle ear infections, noise exposure and age-related hearing loss \cite{WHO}. The World Health Organization estimates that over 1.5 billion people worldwide have hearing loss. This is projected to rise to 2.5 billion by 2050. In the UK, nearly 12 million people -- 1 in 5 -- have hearing loss, with more than 40\% of cases affecting people over 50 years old, and this figure rises to more than 70\% for people over 70 years old \cite{RNID_2022}.

Hearing loss can have a major impact on a person's quality of life, making it difficult to communicate, participate in social activities, and enjoy music. Despite this, only 40\% of people who could benefit from hearing aids actually have them and use them often enough. This is partly because people perceive hearing aids as performing poorly \cite{Kochkin2002, kochkin2010marketrak, knudsen14nielsen,meyer2012factors} or find little benefit in using them \cite{polathearing, tsimpida2020comparison}. Historically, hearing aids have focused on speech communication. However, music listening is also important as it benefits health and well-being. Music is a universal human phenomenon that exists in many contexts and has a powerful impact on our emotions \cite{Mehr2019}. Hearing loss can make it difficult to appreciate music. To take two examples, it can affect the ability of listeners to pick out the lyrics and melody lines, as well as to hear the high frequencies that give the music its richness and detail. As a result, music can sound dull, which can lead to disengagement from music. 




There are several spectro-temporal differences between speech and music that makes hearing aids optimised for speech perform poorly for music \cite{chasin2004hearing}. Although manufacturers have been developing special programs for music listening, the effectiveness has been mixed, with 68\% of users reporting difficulty listening to music through their hearing aids \cite{Greasley2020}. There is a pressing need for better and more inclusive technology to enhance music accessibility to people with hearing loss.

The Cadenza project aims to improve the audio quality of music for people with hearing loss who wear hearing aids, using signal processing and machine learning challenges. These challenges are designed to bring together various research communities to make music more accessible to everyone, taking into consideration the diversity of listeners and making it more inclusive. In the first round (CAD1), we focused on two common scenarios for listening to music: over headphones and in a car in the presence of noise. Firstly, we introduce the general structure and design of CAD1 challenge. Sections \ref{sec:task1} and \ref{sec:task2} describe the specifics of Task 1 and Task 2. We conclude in Section \ref{sec:conclusion}. More details can be found on the challenge website\footnote{\url{https://www.cadenzachallenge.org}}.



\section{Overview of the Challenge Tasks}
\label{sec:methodology}

In the first scenario (Task 1), a person with hearing loss listens to music through headphones without using their hearing aids. In the second scenario (Task 2), the listener is inside a moving car, listening to the music that is coming from the car stereo in the presence of noise, while wearing their hearing aids. Entrants to the challenges are tasked with personalizing the music signals to improve the audio quality.

\begin{figure}[htb]
    \centering
    \includegraphics[width=0.9\linewidth]{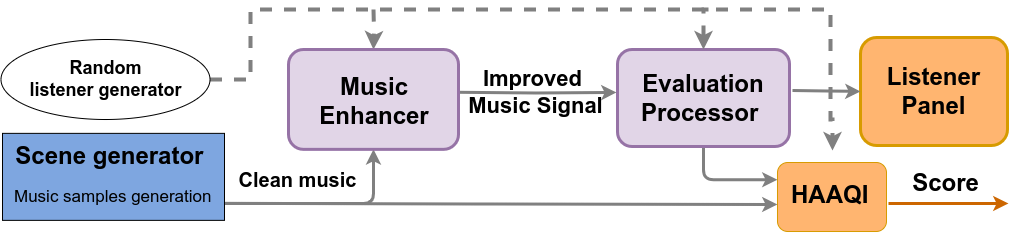}
    \caption{Diagram of the structural design of the first challenge}
    \label{fig:challenge_diagram}
\end{figure}

Figure \ref{fig:challenge_diagram} shows a diagram with the general structure of the challenges. Entrants must develop a \textit{Music Enhancer} that takes in \textit{clean music} and listener characteristics as input. The \textit{Evaluation Processor} then takes the improved music signals, applying any acoustic conditions that are relevant to the task. Finally, the signals are evaluated using the Hearing Aid Audio Quality Index (HAAQI) \cite{Kates2016} and a listener panel.

\subsection{Listener characterisation databases}


Listeners are characterised by bilateral pure-tone audiograms. This give the audible thresholds at standardised frequencies ([250, 500, 1000, 2000, 3000, 4000, 6000, 8000] Hz) as measured by an audiometer \cite{british2004pure}. While a wider frequency range might have been useful for music, we were restricted to this range because these are the standard frequencies that have been tested in the available databases.

For the training (\textit{train}) set, we used the 83 audiograms employed by the 2nd Clarity Enhancement Challenge \cite{CEC2} from the Clarity Project\footnote{\url{https://www.claritychallenge.org}}. These correspond to anonymised examples of real audiograms drawn from the Scottish Section of Hearing Sciences at the University of Nottingham dataset.

For the development (\textit{dev}) set, we selected 50 audiograms from \cite{Von2017}. We first filtered the audiograms to better-ear 4-frequency hearing loss between 20 and 75 dB. We then randomly chose the necessary number of audiograms to maintain the same distribution per band as in the original Clarity dataset. This \textit{dev} set has an equal male-female distribution.

For the evaluation (\textit{eval}) set, we recruited 52 bilateral hearing aid users, with symmetric or asymmetric hearing loss. The listeners were recruited via the University of Leeds. Hearing loss severity was mild for 15 listeners, moderate for 17 listeners, moderately severe for 18 listeners and severe for 2 listeners. In the \textit{train}, \textit{dev} and \textit{eval} sets, hearing loss levels, at each frequency, were limited to 80 dB Hearing Level (HL).

\subsection{Challenge Evaluations}


Both scenarios will be subjected to two evaluation processes. The first is an objective evaluation using HAAQI. This is an intrusive metric in which the processed and reference signals are compared. In the evaluation, the HAAQI function is configured so that the reference signal has an amplification applied to it, so that all frequency bands contribute equally to its loudness. This amplification is the NAL-R hearing aid prescription \cite{Peters2000}. This prescribes the gain to apply based on the individual's audiogram thresholds (in dB HL). This linear amplification improves audibility; there is no dynamic range compression.

The second evaluation consists of a listener panel of 52 listeners (the \textit{eval} listeners) who will rate the audio quality of the music samples. The panel will use a number of scales: \textit{clarity}, \textit{harshness}, \textit{distortion}, \textit{frequency balance}, \textit{overall audio quality}, and \textit{liking}. Overall audio quality captures whether the audio quality is poor or good, and liking is how much the listener liked the specific piece they just listened to. These dimensions have been developed for this purpose, through a sensory evaluation study \cite{Bannister2023}.

Overall audio quality captures whether the audio quality is poor / good, and liking is how much the participant liked the specific piece they just listened to

\section{Design of Task 1}
\label{sec:task1}



This is presented as a demixing-remixing problem. The demixing stage follows the same design as previous music separation challenges \cite{SiSEC18,mitsufuji2021music}. The goal is to decompose stereo music into vocal, drums, bass, and other (VDBO). However, unlike past demixing challenges, we use HAAQI for the evaluation instead of the signal-to-distortion ratio. In the remixing stage, the separated VDBO components allow for personalised remixing for each listener. For example, for some music, the vocals could be amplified to improve the audibility of the lyrics. In Task 1, for Figure \ref{fig:challenge_diagram}, the Evaluation Processor only focuses on computing HAAQI.


We use the MUSDB18-HQ music dataset \cite{musdb18-hq}. This contains 150 tracks where 86 are for \textit{train}, 14 for \textit{dev} and 50 for \textit{eval}. For training data augmentation, entrants are allowed to use the BACH10 \cite{Duan2012}, FMA-Small \cite{fma_dataset} and MedleydB versions 1 \cite{Bittner2014} and 2 \cite{bittner2016}.   


The \textit{eval} set includes 49 tracks, after removing one track where the lyrics might cause offence. For the objective evaluation, 30-second segments of music were selected at random. For the listening panel, 15-second segments were selected at random, ensuring that no explicit language was present.
\section{Design of Task 2}
\label{sec:task2}



The goal is to process music emitted by a car stereo, while accounting for the presence of simulated car noise. However, this is not a denoising problem, as participants do not have access to the exact noise signal. The evaluator processor, for Figure \ref{fig:challenge_diagram}, adds the car acoustic conditions (head-related impulse responses (HRIRs) and simulated car noise) and applies the fixed hearing aid processing algorithms to the enhanced signals before computing the HAAQI score.



The music datasets for task 2 are based on the FMA-Small \cite{fma_dataset} and the MTG Jamendo datasets \cite{Bogdanov2019}. FMA-Small contains 30-second music segments from eight genres; however, we only included the genres `Hip-Hop', `Instrumental', `International', `Pop' and `Rock' as they are the most likely to be found in a car listening environment for our target listeners. We also included samples from the `classical' and `orchestral' genres from the MTG-Jamendo dataset, as people with hearing loss are more likely to be older and listeners to classical music \cite{Bonneville2013}. 


This process resulted in 7000 30-second tracks distributed as follows: 5600 \textit{train}, 700 \textit{dev}, and 700 \textit{eval}. However, as the \textit{eval} samples will be listened to by the listener panel, the \textit{eval} set was reduced to 70 samples by randomly selecting 10 samples per genre. HAAQI will be computed over the whole 30-second segments but, only 15 seconds will be scored by the panel. 


We simulate the different components of the car noise based on a car speed and gear as follows: (i) a mono complex tone corresponding to the engine noise, (ii) two mono signals generated by filtering a white noise by a lowpass filter with a 6-dB/octave slope corresponding to aerodynamic and rolling noise from the left and right side of the car. 

HRIRs were drawn from the eBrIRD - ELOSPHERES database \cite{eBrird}. Each scene contains 74 measurements from -90$^\circ$ to 90$^\circ$ azimuth in 2.5$^\circ$ steps. Anechoic HRIRs are used for the car noise, and in-car HRIRs are used for the enhanced signals.

\section{Baseline Details and Results}

Two baseline systems were proposed for Task 1. The first baseline, `Baseline 1 - Demucs', utilizes the out-of-the-box hybrid-demucs source separation model \cite{defossez2021hybrid} to estimate the VDBO components of the music. This hybrid architecture leverages both time-domain and spectrogram-based approaches. The second baseline, `Baseline 2 - OpenUnmix', employs the OpenUmix source separation model \cite{stoter19}. OpenUnmix is a purely spectrogram-based approach and served as the baseline for the SiSEC 2018 challenge \cite{SiSEC18}.

For Task 2, a single baseline was proposed, which applies a level constraint to the mixture at the hearing aid microphones to prevent clipping caused by the NAL-R hearing aid amplification.

Table \ref{tab:baseline_results} shows the baseline HAAQI scores for Task 1 and Task 2. For Task 1, the results correspond to the mean and standard deviation (std) of all eight left and right VDBO signals. In Task 2, the scores represent the mean and std of the scores per genre.

\begin{table}  
    \caption{Baseline HAAQI scores for Task 1 and Task 2 of the First Cadenza Challenge.}
    \label{tab:baseline_results}
    \begin{tabular}{l|l|c|c}
        \toprule
            \textbf{Task} & \textbf{System} & \textbf{mean HAAQI} & \textbf{std HAAQI} \\ 
        \midrule
            \multirow{2}{*}{Task 1}  & Baseline 1 - Demucs & 0.2548 & 0.0403 \\
            & Baseline 2 - OpenUnmix & 0.2253 & 0.0290  \\
        \midrule
            Task 2  & Baseline & 0.1256 & 0.0321  \\ 
        \bottomrule
    \end{tabular}
\end{table}
\section{Conclusion and Future Work}
\label{sec:conclusion}


The Cadenza project aims to improve the audio quality of music for those who have a hearing loss. It consists of two common music listening scenarios; Task 1, listening to music over headphones and, in Task 2, listening to music in a car. While Task 1 presents a demixing-remixing problem, Task 2 presents a near-end music enhancement problem. Both scenarios will be evaluated using HAAQI for objective evaluation and a listener panel for subjective evaluation in autumn/winter 2023. The project team will run an online workshop in December 2023 where entrants will outline their approaches, and the results of the evaluations will be presented. Currently, we are running an ICASSP 2024 Grand Challenge and the CAD2 challenge will launch in 2024.

\begin{acknowledgments}
 The Cadenza project is supported by the Engineering and Physical Sciences Research Council (EPSRC) [grant number: EP/W019434/1]. \\We thank our partners: BBC, Google, Logitech, RNID, Sonova, Universit\"{a}t Oldenburg. 
\end{acknowledgments}

\bibliography{references}


\end{document}